\documentclass[%
 preprint,
 superscriptaddress,
 amsmath,amssymb,
 aps,
prb,
 floatfix,
]{revtex4-1}

\usepackage{graphicx}
\usepackage{dcolumn}
\usepackage{bm}
\usepackage{hyperref}
\usepackage[modulo]{lineno}
\usepackage{setspace}

\usepackage{xcolor}

\begin{document}
\setstretch{1.25}

\title{Patterns and driving forces of dimensionality-dependent charge density waves in 2\textit{H}-type transition metal dichalcogenides}

\author{Dongjing~Lin}
\author{Shichao~Li}
\affiliation{National Laboratory of Solid State Microstructures and Department of Physics, Nanjing University, Nanjing 210093, China.}

\author{Jinsheng~Wen}
\affiliation{National Laboratory of Solid State Microstructures and Department of Physics, Nanjing University, Nanjing 210093, China.}
\affiliation{Collaborative Innovation Center of Advanced Microstructures, Nanjing University, Nanjing 210093, China.}

\author{Helmuth~Berger}
\author{L\'{a}szl\'{o} Forr\'{o}}
\affiliation{Institute of Condensed Matter Physics,
\'{E}cole Polytechnique Fed\'{e}ral\'{e} de Lausanne, 1015 Lausanne, Switzerland.}

\author{Huibin~Zhou}
\author{Shuang~Jia}
\affiliation{International Center for Quantum Materials, School of Physics, Peking University, Beijing 100871, China.}
\affiliation{Collaborative Innovation Center of Quantum Matter, Beijing 100871, China.}

\author{Takashi~Taniguchi}
\author{Kenji~Watanabe}
\affiliation{National Institute for Materials Science, 1-1 Namiki, Tsukuba 305-0044, Japan.}

\author{Xiaoxiang~Xi}
\email{xxi@nju.edu.cn}
\affiliation{National Laboratory of Solid State Microstructures and Department of Physics, Nanjing University, Nanjing 210093, China.}
\affiliation{Collaborative Innovation Center of Advanced Microstructures, Nanjing University, Nanjing 210093, China.}

\author{Mohammad~Saeed~Bahramy}
\email{bahramy@ap.t.u-tokyo.ac.jp}
\affiliation{Quantum-Phase Electronics Center (QPEC) and Department of Applied Physics, The University of Tokyo, Tokyo 113-8656, Japan.}
\affiliation{RIKEN Center for Emergent Matter Science (CEMS), Wako 351-0198, Japan.}

\maketitle

\noindent\textbf{Two-dimensional (2D) materials have become a fertile playground for the exploration and manipulation of novel collective electronic states. Recent experiments have unveiled a variety of robust 2D orders in highly-crystalline materials ranging from magnetism\cite{Gong2017,Huang2017} to ferroelectricity\cite{Chang2016,Liu2016,Fei2018} and from superconductivity\cite{Saito2016} to charge density wave (CDW) instability\cite{Ugeda2016,Xi2015}. The latter, in particular, appears in diverse patterns even within the same family of materials with isoelectronic species. Furthermore, how they evolve with dimensionality has so far remained elusive. Here we propose a general framework that provides a unified picture of CDW ordering in the 2H polytype of four isoelectronic transition metal dichalcogenides 2\textit{H-MX}$_2$ (\textit{M}$=$Nb, Ta and \textit{X}$=$S, Se). We first show experimentally that whilst NbSe$_2$ exhibits a strongly enhanced CDW order in the 2D limit, the opposite trend exists for TaSe$_2$ and TaS$_2$, with CDW being entirely absent in NbS$_2$ from its bulk to the monolayer. Such distinct behaviours are then demonstrated to be the result of a subtle, yet profound, competition between three factors: ionic charge transfer, electron-phonon coupling, and the spreading extension of the electronic wave functions. Despite its simplicity, our approach can, in essence, be applied to other quasi-2D materials to account for their CDW response at different thicknesses, thereby shedding new light on this intriguing quantum phenomenon and its underlying mechanisms.}

CDWs are the ground state of some low-dimensional metals, featuring simultaneous periodic modulation of the mobile charge density and distortion of the crystal lattice\cite{Gruner2000}. Their mechanism has been under close scrutiny for decades\cite{Wilson2001,Rossnagel2011} and their relation with superconductivity bears important implications regarding the nature of the superconducting states\cite{Fradkin2015,Chen2016}. The advent of 2D materials\cite{Novoselov2005} ushers in a new era of exploring CDWs in confined dimensions\cite{Calandra2009,Ge2012}. Mechanical exfoliation\cite{Xi2015,Hajiyev2013,Yu2015}, molecular-beam epitaxy (MBE)\cite{Ugeda2016,Chen2015,Sugawara2016,Ryu2018}, and chemical vapour deposition\cite{Fu2016,Wang2018} have produced a plethora of 2D materials exhibiting CDWs. Dimensionality reduction was shown to enhance the CDW order in some of them\cite{Xi2015,Chen2015,Ryu2018,Wang2018} while suppress it in others\cite{Ugeda2016,Yu2015,Fu2016,Yang2018}. Even for the same compound\cite{Ugeda2016,Xi2015,Chen2015,Sugawara2016}, results from different studies are scattered. For better understanding and control of the CDW states, it is thus imperative to determine the key factors governing the intrinsic effects of dimensionality on CDWs.

In this work, we systematically study the CDWs in group-V transition metal dichalcogenides 2\textit{H-MX}$_2$ (\textit{M}$=$Nb, Ta and \textit{X}$=$S, Se). We introduce a chemical-bonding framework that can explain the distinct CDW properties observed in these systems and materials alike. The prototypes chosen here all crystalize in the so-called 2\textit{Ha} structure\cite{Wilson2001} (Figure 1a). The transition metal and chalcogen atoms form an \textit{X-M-X} sandwich in the trigonal prismatic coordination, henceforce called a monolayer. The unit cell for the bulk crystals consists of two such monolayers rotated 180$^{\circ}$ from each other and stacked with the transition metal atoms aligned along the $c$ axis. Common to all these compounds is a mismatch of valency between the central \textit{M} ion ($3+$) and the neighbouring chalcogens (each in $2-$ state). As such, they all show a metallic behaviour with a mixed ionic-covalent character. Figure 1a illustrates three major pathways for charge distribution in such a metallic environment: (1) an incomplete ionic charge transfer between the transition metal cation and its adjacent chalcogen anions within a monolayer (denoted as $\Delta Q_I$ and determined by their electronegativity difference, see Supplementary Section 1) forming the ionic \textit{M}-\textit{X} bonds; (2) intralayer axial and planner $\sigma$-electron hoppings between the chalcogen \textit{p} orbitals, both forming covalent \textit{X}-\textit{X} bonds; (3) interlayer $\sigma$-hoppings across the van der Waals gap.

The contribution of each channel to CDWs can be assessed according to whether the formed chemical bonds make the lattice ‘harder’ or ‘softer’. In principle, a harder lattice is less susceptible to distortions, hence less amenable to CDW formation. This argument suggests the reduced tendency of CDW transitions in materials with sizable ionic charge transfer, as stronger ionic bonds make the lattice more rigid. However, the spatial extent of the ionic charge transfer crucially influences the strength of the ionic bonds. For example, the electronic bandwidth of the topmost valence band in monolayer TaS$_2$ is nearly 140\% of that for monolayer NbSe$_2$ (Figure 1b and 1c; for a full comparison also see Supplementary Figure S2), thereby enabling a more extended charge distribution (as compared in Figure 1d and 1e). This is ascribed to the large spin-orbit coupling of heavy tantalum atoms as well as the increased screening of the Coulomb potential from their nuclei due to their additional shells of semi-core and core states. Because of the extended charge distribution, despite the substantial total ionic charge transfer for TaS$_2$ shown in Figure 1f, multiple bonds are involved in this sharing. Hence, its net effect on the lattice is counteracted in favour of CDW transition. In contrast, intralayer covalency, more pronounced in Nb-based compounds, is unfavourable for CDWs, as it pins the layers to each other, and thus, hardens the whole lattice. Interlayer $\sigma$-hopping works in unison with intralayer covalent bonding to account for the dimensionality-dependent CDWs in these compounds. This complexity in charge distribution obviously hinders the predictability of CDW phase transition in general and its dependence on dimensionality, specifically. However, as we discuss later, behind such chaos there appears to be an order that can be rationalized purely based on the chemistry of metals. Before that, let us first present the characteristic CDW features that we have observed experimentally in these compounds.

High-quality single crystals were chosen for our experiments. Their temperature-dependent resistance data in Figure 2a show typical metallic behaviour, in stark contrast to the metal-insulator transition expected for Pierels instability. CDW transitions manifest as a weak hump at $\sim$30 K for NbSe$_2$ and as clear kinks at $\sim$80 K and $\sim$120 K for TaS$_2$ and TaSe$_2$, respectively. NbS$_2$ lacks CDW order and thus does not show an anomaly in the resistance. These temperatures for the incommensurate CDW transition ($T_{\mathrm{CDW}}$) as well as the superconducting transitions at a lower temperature are consistent with those established in the literature\cite{Naito1982}. Atomically thin samples were mechanically exfoliated from the bulk crystals, with special care taken to minimize sample degradation (see Methods).

We used Raman scattering to obtain both CDW signatures and phonon information\cite{Sugai1985}. Figure 2b shows room temperature Raman spectra of the bulk compounds, collected in the collinear (XX) and cross (XY) polarization configurations. In the backscattering geometry of our experiment, the former detects both $A_{1g}$ and $E_{2g}$ symmetry while the latter only couples to $E_{2g}$\cite{Xi2015}. Four common features are noted in the data: (1) a rigid layer mode with $E_{2g}$ symmetry (labelled as shear mode), which corresponds to interlayer shearing vibration and distinguishes the 2\textit{H} from 1\textit{T} and 3\textit{R} polytypes; (2) an $A_{1g}$ phonon mode, which only involves the chalcogen atoms vibrating against each other along the $c$ axis; (3) an $E_{2g}$ phonon mode, which involves the transition metal and chalcogen atoms vibrating along opposite directions within the layer plane; (4) a broad two-phonon scattering peak commonly assigned as due to the CDW soft phonons\cite{Sugai1985,Hill2019,Joshi2019}. The vibration patterns for the $A_{1g}$ and $E_{2g}$ modes suggest that the ionic bonds dominate their eigenfrequencies. As shown in Figure 2c, both the $A_{1g}$ and $E_{2g}$ modes are found to have much higher frequencies in S-based 2\textit{H-MX}$_2$ compounds than in Se-based ones. Sulphur due to its higher electronegativity can gain more charge from the \textit{M} site, thereby making the resulting \textit{M}-S bond harder than its \textit{M}-Se counterpart in 2\textit{H-M}Se$_2$. This is consistent with the trend of ionic charge transfer shown in Figure 1f. 

Figure 2d compares the 300 K and 4 K spectra, with new modes emerging at low temperature in all compounds except NbS$_2$. These are the amplitude modes or zone-folded modes unique to the CDW phase\cite{Xi2015,Hill2019,Joshi2019}. The former features characteristic softening and broadening upon approaching $T_{\mathrm{CDW}}$ from below, while the latter shows only a minute change of the frequency. Figure 3b, d, and f show the layer number dependence of these modes at 4 K for NbSe$_2$, TaSe$_2$, and TaS$_2$, respectively. We focus on the amplitude modes, which are the most pronounced features in the spectra below 100 cm$^{-1}$. NbSe$_2$ exhibits a single broad peak that shifts from below 50 cm$^{-1}$ in the bulk to 73 cm$^{-1}$ in the monolayer, while the intensity is less affected. In contrast, both TaSe$_2$ and TaS$_2$ exhibit multiple sharper amplitude modes below 100 cm$^{-1}$. Each peak shows a weak layer number dependence for its frequency, while the intensity dramatically diminishes in atomic layers. The blueshift of the mode frequency (suppression of the mode intensity) correlates well with enhanced (reduced) $T_{\mathrm{CDW}}$ in atomically thin NbSe$_2$ (TaSe$_2$ and TaS$_2$), as will be detailed below. The completely different behaviours of the amplitude modes in the Nb- and Ta-based compounds call for further investigations.

We do not observe any new modes at low temperature in NbS$_2$ down to the monolayer (Supplementary Figure S3), therefore concluding the entire absence of CDW in this compound. Previous work has shown latent CDW in bulk NbS$_2$ due to the strong anharmonicity of the lattice potential\cite{Leroux2012}. Our temperature-dependent study of the $A_{1g}$ mode frequency shows that NbS$_2$ indeed exhibits the largest degree of lattice anharmonicity among the four compounds (Supplementary Section 3.2). Our calculation shown in Figure 1f suggests enhanced ionic bonding upon reducing the layer number in NbS$_2$. CDWs, therefore, do not emerge in atomically thin samples, although a recent study predicts otherwise\cite{Bianco2019}.

Temperature and layer-number dependent Raman measurements show systematic evolution of the amplitude modes in NbSe$_2$, TaS$_2$, and TaSe$_2$ (Supplementary Figure S3). To facilitate comparison of the mode intensity among different samples, we transform the raw spectra $I$ to $I/I_0-1$, where $I_0$ is a high-temperature spectrum far above $T_{\mathrm{CDW}}$. The normalized temperature-dependent Raman scattering intensity maps for samples of various thickness are shown in Figure 3a, c and e. The amplitude modes are accentuated in red, which diminish into the background upon increasing temperature. For NbSe$_2$, we clearly identify a strong enhancement of $T_{\mathrm{CDW}}$ in the bilayer and monolayer, consistent with a previous Raman study using a different excitation laser energy\cite{Xi2015}. For TaSe$_2$ and TaS$_2$, although the mode intensity is significantly weakened in the atomically thin samples, the transition temperature appears similar to that of the bulk.

Figure 4a summarises the layer number dependence of $T_{\mathrm{CDW}}$ for each compound, estimated from the temperature dependence of the amplitude mode intensity, integrated with respect to a featureless background (Supplementary Section 4). The figure clearly shows enhanced (suppressed) $T_{\mathrm{CDW}}$ in NbSe$_2$ (TaSe$_2$ and TaS$_2$) when approaching the monolayer limit. The trends are further confirmed by analysing the amplitude mode frequency and the zone-folded mode intensity (Supplementary Sections 5--7). NbS$_2$ do not show Raman signature of CDWs for the measured bulk, bilayer, and monolayer samples, hence their vanishing $T_{\mathrm{CDW}}$.

The disparate thickness-dependent $T_{\mathrm{CDW}}$ observed for the isostructural and isoelectronic compounds in the same material family is highly unusual. Since the significant two-phonon scattering peak in these compounds originates from the longitudinal acoustic phonon branch exhibiting Kohn anomaly\cite{Sugai1985,Hill2019,Joshi2019}, and because its frequency and temperature variation show rather weak thickness dependence (Supplementary Section 8), we infer a nearly thickness-independent CDW wavevector for NbSe$_2$, TaSe$_2$, and TaS$_2$. The thickness dependence of $T_{\mathrm{CDW}}$ is, therefore, not related to different forms of superlattice. Instead, it appears to originate from the intrinsic chemical properties of the ingredients, constituting these materials. First of all, the existence of CDWs in monolayer NbSe$_2$, TaSe$_2$, and TaS$_2$ is permitted by their incomplete ionic bonding. As we explained earlier, this is primarily due to a mismatch between the valence state of \textit{M} cations and \textit{X} anions. Furthermore, the occurring ionic charge transfer is either small and spatially highly localized as in the case of NbSe$_2$ or significant but spatially extended as in the case of TaSe$_2$ and TaS$_2$. In going from a monolayer to bulk, the nature of bonding between the layers and their effect on the spread of the wave functions defines the fate of the CDW in each compound. If the interlayer bonding enhances the rigidity of each layer, $T_{\mathrm{CDW}}$ is expected to show declining behaviour, as the thickness grows. On the other hand, if it weakens the initial bonding, the result will manifest as an enhancement of CDW instability. 

The parameter which quantifies either of these trends is the intralayer covalency, $\Delta Q_{\sigma}$. We define this parameter as the differential charge at the chalcogen sites upon migration from a monolayer to a multilayer system. A positive $\Delta Q_{\sigma}$ means that the interlayer hopping has helped each chalcogen to gain more charge, thereby making the whole entity a ``harder lattice''. Needless to say, a negative $\Delta Q_{\sigma}$ means the added layers have led to a ``softer lattice''. Using atomic-orbital-like Wannier functions (see Methods), we have calculated $\Delta Q_{\sigma}$ for all four compounds in the bilayer and bulk configurations (Figure 1g). As can be seen, both bilayer and bulk NbSe$_2$ show a significant gain in $\Delta Q_{\sigma}$, implying that their crystal structure is relatively harder than that of monolayer NbSe$_2$, and thus, less prone to CDW instability. We attribute this to the localized nature of wave functions in NbSe$_2$, mainly made up by the axial orbitals Se-$p_z$ and Nb-$d_{z^2}$ (see Figure 1d). In a bilayer or bulk NbSe$_2$, this accordingly allows a direct $\sigma$-type interlayer hopping between these orbitals, making $\Delta Q_{\sigma}$ positive. More importantly, this additional bonding acts against the $A_{1g}$ phonon mode, resulting in a decrease in the strength of electron-phonon coupling constant $\lambda$ as depicted in Supplementary Figure S14. It then becomes clear why $T_{\mathrm{CDW}}$ in NbSe$_2$ shows a negative thickness dependence.

TaS$_2$ and TaSe$_2$ behave differently. As shown in Figure 1g, both compounds have a negative $\Delta Q_{\sigma}$ with lower values found for their bulk phases. Given their extended wave functions, the only channel allowing the interlayer coupling is $\sigma$-hopping between the chalcogen $p_z$ orbitals of neighbouring layers. As such, the initial charge residing on the chalcogen sites is partly shifted between the layers. This, in turn, leads to a softening of the original metal-chalcogen bonding and correspondingly an enhancement of CDW instability, as observed experimentally. Interestingly, the calculated $\Delta Q_{\sigma}$ for TaS$_2$ turns out to be more sensitive to the number of layers as compared with that obtained for TaSe$_2$. Logically, one then expects a higher $T_{\mathrm{CDW}}$ in, for example, bulk TaS$_2$ than in bulk TaSe$_2$. In reality, however, TaSe$_2$ exhibits a higher $T_{\mathrm{CDW}}$ than TaS$_2$ regardless of the number of the layers. To address this apparent inconsistency, we need to include another critical parameter in our consideration. That parameter is electron-phonon coupling $\lambda$. A large $\lambda$ can mediate CDW more effectively than a small $\lambda$. However, this requires a proper coupling between the phonon mode eigenvalues and electronic eigenvalues. If the phonon modes, responsible for CDW instability, have relatively high frequencies, such a coupling is not well established and, hence, a relatively small $\lambda$ is achieved. This is what occurs in TaS$_2$. As discussed earlier, for both $A_{1g}$ and $E_{2g}$ modes, the corresponding phonon frequencies are higher in TaS$_2$ than in TaSe$_2$ (See Figure 2c), suggesting that the former should have a smaller $\lambda$ than the latter. Our first-principles calculations for monolayer 2\textit{H-MX}$_2$ confirm this indeed. In fact, as shown in Figure 4b, we find that $\lambda$ is always smaller in an S-based compound when compared to its Se-based counterpart. These calculated results agree well with the available $\lambda$ values reported for NbSe$_2$\cite{Zheng2019}, TaS$_2$ and TaSe$_2$\cite{Hinsche2017}.

In a broader context, we can again attribute this tendency to the ionic charge transfer $\Delta Q_I$ we discussed earlier, implying a trade-off between $\Delta Q_I$ and $\lambda$. In addition to these two parameters, the spatial extension of the wave functions (represented here as $1/\langle r^2\rangle$) plays an important role. The more localized are the wave functions, the less effective become both $\Delta Q_I$ and $\lambda$. This is because it enhances the electron-electron correlation, demobilizing the carriers and so preventing the formation of any superlattice charge ordering. As a result, if $\lambda$ is not large enough, CDW remains forbidden. This then solves the last piece of the puzzle, i.e. why NbS$_2$ is so robust against CDW instability. We note that van Loon et al. have already given a detailed account for the competing nature of electron-phonon coupling versus the short- and long-range electron Coulomb interaction, and its prominent role in the absence of CDW instability in NbS$_2$\cite{Loon2018}. Interestingly, they suggest that the interplay between these parameters strongly enhances both charge and spin susceptibilities in NbS$_2$, meaning that it is at the verge of collapsing to a CDW or even a spin density wave phase, if sufficient perturbations are introduced (for example via local magnetic impurities)\cite{Loon2018}. As such, NbS$_2$ appears to be an ideal candidate for exploring thermal and quantum fluctuations. 

Altogether, the three parameters (1) ionic charge transfer, (2) electron-phonon coupling, and (3) the spatial extension of the electronic wave functions are the key components defining the fate of CDW ordering and its thickness dependence in this and potentially other layered materials. Thus, they can be used to create a unified phase diagram describing such instabilities in low-dimensional limits. We have schematically illustrated such a phase diagram in Figure 4c. As can be seen, the interplay between these three parameters is expected to form an abyss-like shape, deepening at regions with low $\Delta Q_I$ and $\lambda$ and high $1/\langle r^2\rangle$. Above the surface of the resulting terrain is a parameter-space within which the CDW is forbidden and below it is where CDW is allowed to emerge. We also have shown the schematic location of each of the 2\textit{H-MX}$_2$ compounds, studied here. Obviously, NbS$_2$ is the only member of this family appearing in a region at which CDW is prohibited.   

Lastly, we address some controversies in the studies of 2D CDWs in NbSe$_2$, TaSe$_2$, and TaS$_2$. For NbSe$_2$, a strongly enhanced $T_{\mathrm{CDW}}$ from $\sim$33 K in the bulk to $\sim$145 K in the monolayer was previously reported in mechanically exfoliated samples on sapphire substrates\cite{Xi2015} and confirmed here. This is in stark contrast to the almost unchanged $T_{\mathrm{CDW}}$ in MBE-grown monolayer NbSe$_2$ on bilayer graphene\cite{Ugeda2016}. For TaSe$_2$, we found a rather weak suppression of $T_{\mathrm{CDW}}$ in the exfoliated monolayer, while MBE-grown monolayer TaSe$_2$ on bilayer graphene shows slightly enhanced $T_{\mathrm{CDW}}$ with respect to its bulk value\cite{Ryu2018}. The exfoliated samples are expected to exhibit intrinsic CDW properties, verified by control experiments on suspended thin flakes (Supplementary Section 10). The MBE-grown samples may be affected by charge transfer from the underlying graphene substrate. For monolayer NbSe$_2$, such charge transfer can increase the intralayer covalency and suppress the strongly enhanced $T_{\mathrm{CDW}}$. For monolayer TaSe$_2$, due to its extended charge distribution, charge transfer tends to accumulate between TaSe$_2$ and graphene, so that the intralayer covalency is reduced; CDW is therefore enhanced. CDWs were found to be absent in epitaxial monolayer TaS$_2$ on gold substrate\cite{Sanders2016}, clearly deviating from the intrinsic properties found in the exfoliated TaS$_2$ studied here. This is again due to the charge transfer from the substrate.

In conclusion, our chemical-bonding framework provides an intuitive guide for boosting $T_{\mathrm{CDW}}$, which is essential for developing CDW-based devices for applications. It may also be applied to explain the distinct dimensionality effects in the growing family 2D CDW materials\cite{Hossain2017}. Future work should address the connection of the current framework with the existing theories for CDWs, for instance, Fermi-surface nesting and momentum-dependent electron-phonon coupling\cite{Rossnagel2011,Johannes2008}.

\bibliographystyle{naturemag}
\bibliography{reference}

\vspace{3mm}
\noindent
\textbf{Methods}

\noindent
\textbf{Sample preparation.} Bulk single crystals were synthesized by the chemical vapour transport method. Atomically thin flakes were mechanically exfoliated from the bulk crystals on silicone elastomer polydimethylsiloxane stamps and transferred on sapphire substrates for Raman study. To minimize sample degradation, we prepared them in a glove box filled with nitrogen gas, followed by encapsulation with thin h-BN. The flake thickness was determined by the shear mode frequency in the Raman data (Supplementary Section 3.1).

\noindent
\textbf{Characterizations.} Temperature dependent Raman scattering measurements were performed using a home-built confocal optical setup, consisting mainly of a Montana Instruments Cryostation and a Princeton Instruments grating spectrograph quipped with a liquid-nitrogen-cooled charge coupled device. Beam from a diode-pumped 532 nm laser was focused on the sample using a 40$\times$ microscope objective. The backscattered light was collected using the same objective followed by a couple of Bragg notch filters, achieving a minimum cutoff of 15 cm$^{-1}$ in the collinear polarization configuration. The sample chamber was evacuated to high vacuum better than $10^{-4}$ Pa throughout the experiment. Rapid temperature control was achieved using an Agile Temperature Sample Mount. To minimize laser heating, the incident power was kept below 0.1 mW for bulk samples and below 0.3 mW for thin flakes, and anti-Stokes lines were confirmed to be absent at the base temperature. Four-probe resistance measurements on the bulk crystals were conducted in an Oxford Instruments TeslatronPT system using the standard lock-in method.

\noindent
\textbf{Calculations.} The electronic structure of 2\textit{H-MX}$_2$ monolayers was calculated within density functional theory (DFT) using Perdew-Burke-Ernzerhof exchange-correlation functional\cite{Perdew1996} as implemented in Quantum Espresso program package\cite{Giannozzi2017,Giannozzi2009,QE1}. We used the norm-conserving pseudo-potentials\cite{QE2} and plane-wave basis set with cut-off energy of 75 Ry. The relativistic effects, including spin-orbit coupling, were fully considered. The Brillouin zone was sampled by a 24$\times$24$\times$1 $k$-mesh. The ionic charge transfer and hopping parameters were obtained by constructing a 22-band tight-binding model by downfolding the DFT Hamiltonian using maximally localized Wannier functions\cite{Souza2001,MOstofi2008}. The \textit{M}-3$d$ and \textit{X-p} atomic orbitals were taken as projection centres.

For the calculation of phonon modes and the corresponding electron-phonon coupling parameters, we first fully optimized both the lattice parameters and atomic positions until the magnitude of the force on each ionic site was less than $10^{-5}$ Ry/Bohr and the total energy is converged below $10^{-10}$ Ry. To avoid the negative phonon modes, resulting in an overestimation of $\lambda$, the Methfessel-Paxton smearing scheme with a relatively large broadening parameter $\sigma=0.03$ Ry was used, as suggested in Ref.\cite{Zheng2019}. The dynamical matrix was then calculated based on density-functional perturbation theory employing an 8$\times$8$\times$1 $q$-mesh. Finally, we computed $\lambda$ using the interpolation scheme implemented in Quantum Espresso\cite{QE1}.     
   
\vspace{3mm}
\noindent
\textbf{Acknowledgements}\\
This work was support by the National Natural Science Foundation of China (Grant Nos. 11774151, 11822405 and 11674157), the National Key Research and Development Program of China (Grant Nos. 2018YFA0307000 and 2017YFA0303201), the Natural Science Foundation of Jiangsu Province (Grant No. BK20180006) and Japan Science and Technology Agency (CREST, JST, Grant No. JPMJCR16F1). The work in Lausanne was supported by the Swiss National Science Foundation. The crystal growth in Peking University was supported by the National Natural Science Foundation of China (Grant Nos. U1832214 and 11774007) and the National Key Research and Development Program of China (2018YFA0305601). Growth of hexagonal boron nitride crystals was supported by the Elemental Strategy Initiative conducted by the MEXT, Japan and the CREST (JPMJCR15F3), JST.

\vspace{3mm}
\noindent
\textbf{Author contributions}\\
X.X. and M.S.B. conceived the project. D.L. and X.X. performed the experiments. S.L., J.W., H.B., L.F., H.Z. and S.J. contributed the transition metal dichalcogenides crystals. T.T. and K.W. grew the h-BN crystals. D.L. and X.X. analysed the experimental data. M.S.B. performed the first-principles calculations. X.X. and M.S.B. interpreted the results and co-wrote the paper, with comments from all authors.
   
\newpage
   
\begin{figure*}[t]
\includegraphics[scale=0.275]{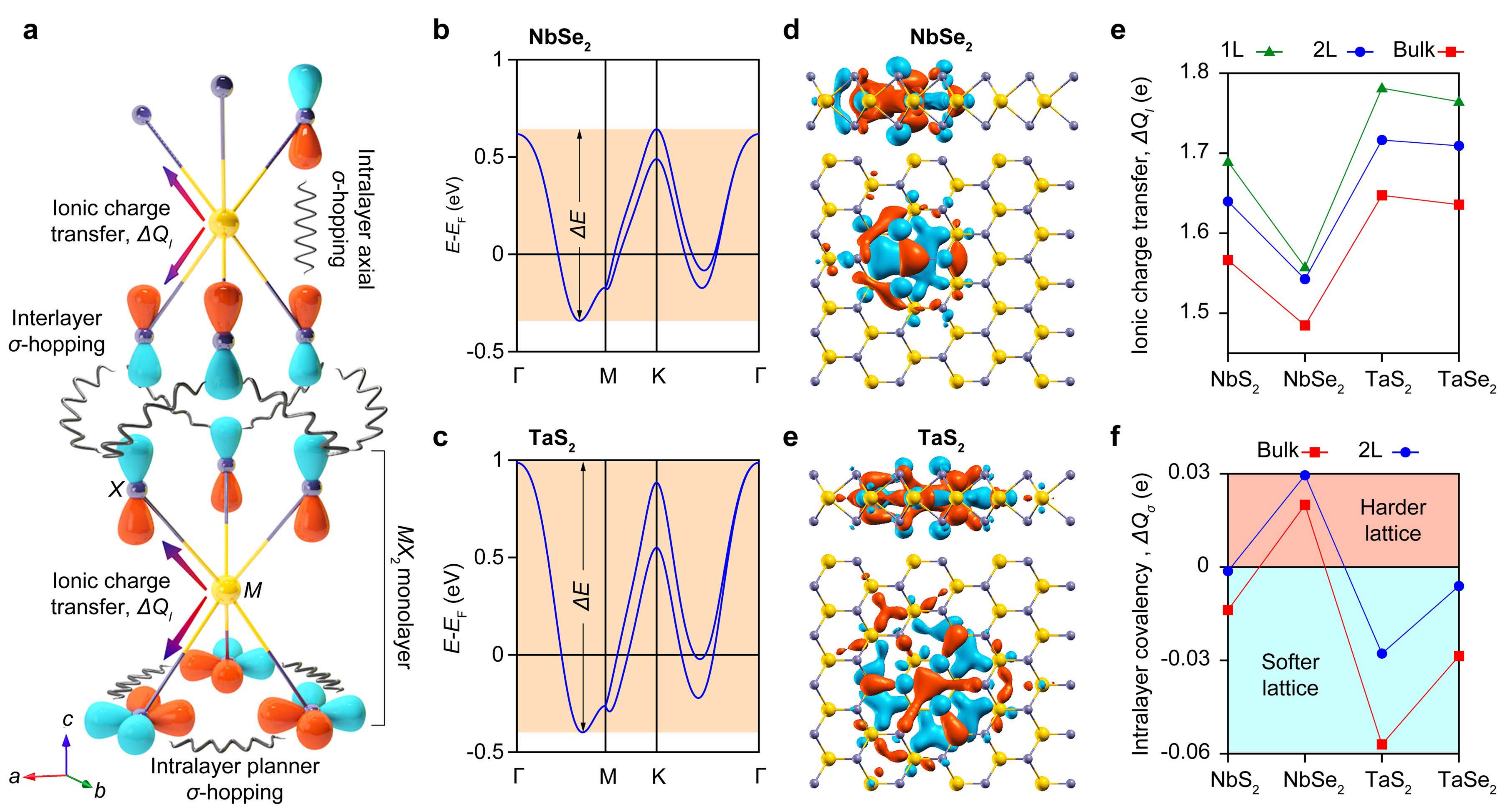}   
\caption{\textbf{Chemical bonding mechanism in 2\textit{H-MX}$_2$ (\textit{M}$=$Nb, Ta and \textit{X}$=$S, Se).} \textbf{a}, Schematic illustration of the charge transfer channels, including intralayer ionic charge transfer $\Delta Q_I$, intralayer covalent bonding (axial $\sigma$-hopping and planner $\sigma$-hopping), and interlayer $\sigma$-hopping. \textbf{b} and \textbf{c}, Electronic band structure for monolayer NbSe$_2$ and TaS$_2$ from first-principles calculations. $\Delta E$ denotes the bandwidth. \textbf{d} and \textbf{e}, calculated charge distribution in monolayer NbSe$_2$ and TaS$_2$, viewed along the in-plane and out-of-plane directions. Orange and blue represents positive and negative signs of the electronic wave functions, respectively. \textbf{f} and \textbf{g}, Layer number dependence of the ionic charge transfer $\Delta Q_I$ and the change of the intralayer covalency $\Delta Q_{\sigma}$ upon increasing the layer number from a monolayer for all compounds.} 
\label{Fig1}
\end{figure*}

\begin{figure*}
\includegraphics[scale=0.47]{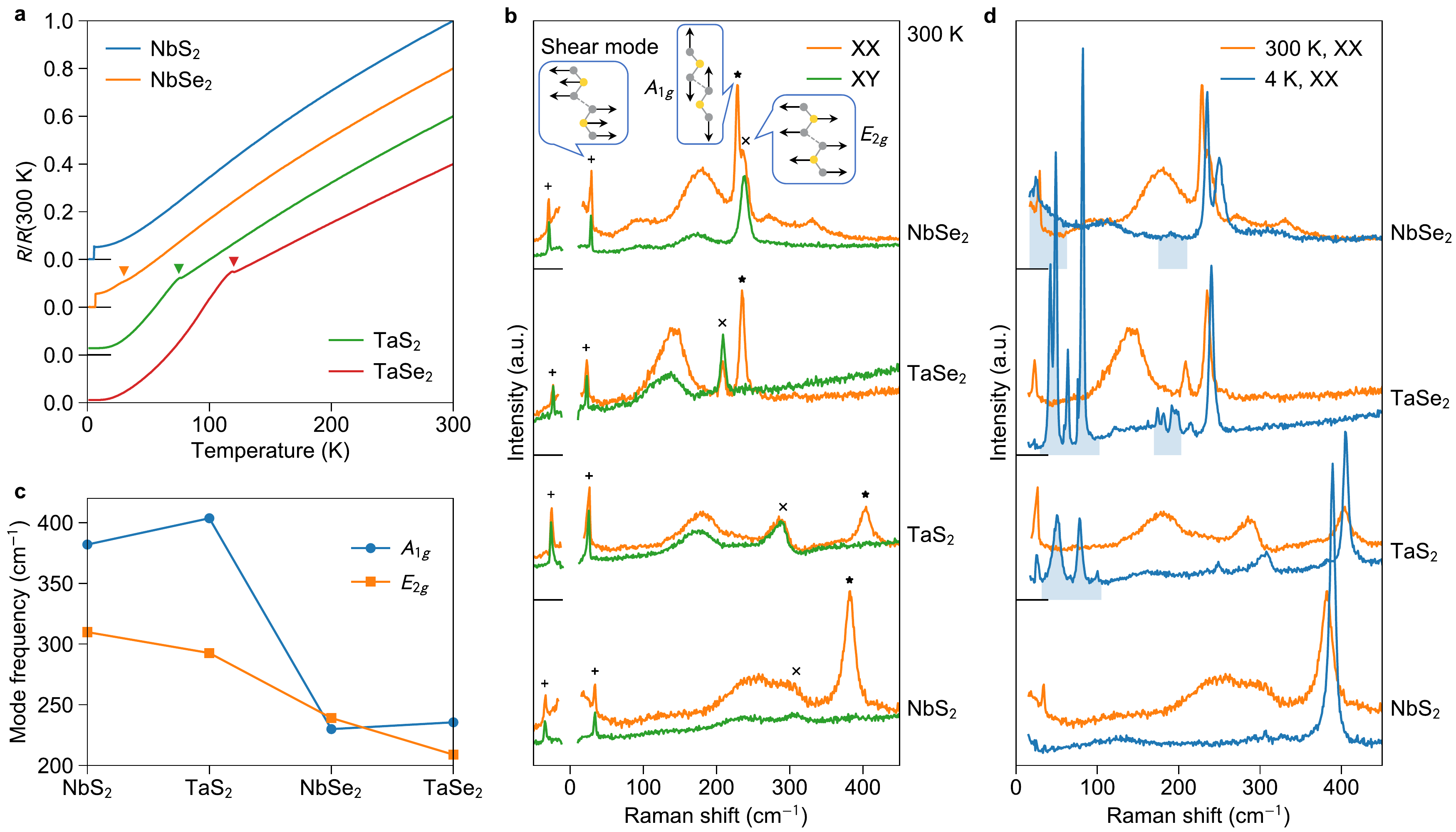}   
\caption{\textbf{Characterizations of phonon modes in high-quality bulk 2\textit{H-MX}$_2$ (\textit{M}$=$Nb, Ta and \textit{X}$=$S, Se) samples.} \textbf{a}, Temperature dependence of the electrical resistance for all compounds, normalized to their respective values at 300 K. The triangles mark the anomalies due to CDW transitions. \textbf{b}, Raman spectra for all compounds collected in the collinear (XX) and cross (XY) polarization configurations at 300 K. The first-order phonon scattering peaks are indicated by $+$ for the shear modes (Stokes and anti-Stokes), $\times$ for the $E_{2g}$ modes, and $\star$ for the $A_{1g}$ modes. The corresponding displacement patterns are illustrated in the balloons. \textbf{c}, Compound dependence of the $A_{1g}$ and $E_{2g}$ mode frequencies analyzed by peak fitting of the data in \textbf{b}. \textbf{d}, Comparison of the Raman spectra at 300 K and 4 K for the collinear polarization configuration for all compounds. CDW-induced modes are highlighted by the shaded regions. Data in \textbf{a}, \textbf{b}, and \textbf{d} are shifted vertically for clarity, with the new origins located at the crossings of the left axes and the long horizontal bars.} 
\label{Fig2}
\end{figure*}

\begin{figure*}
\includegraphics[scale=0.47]{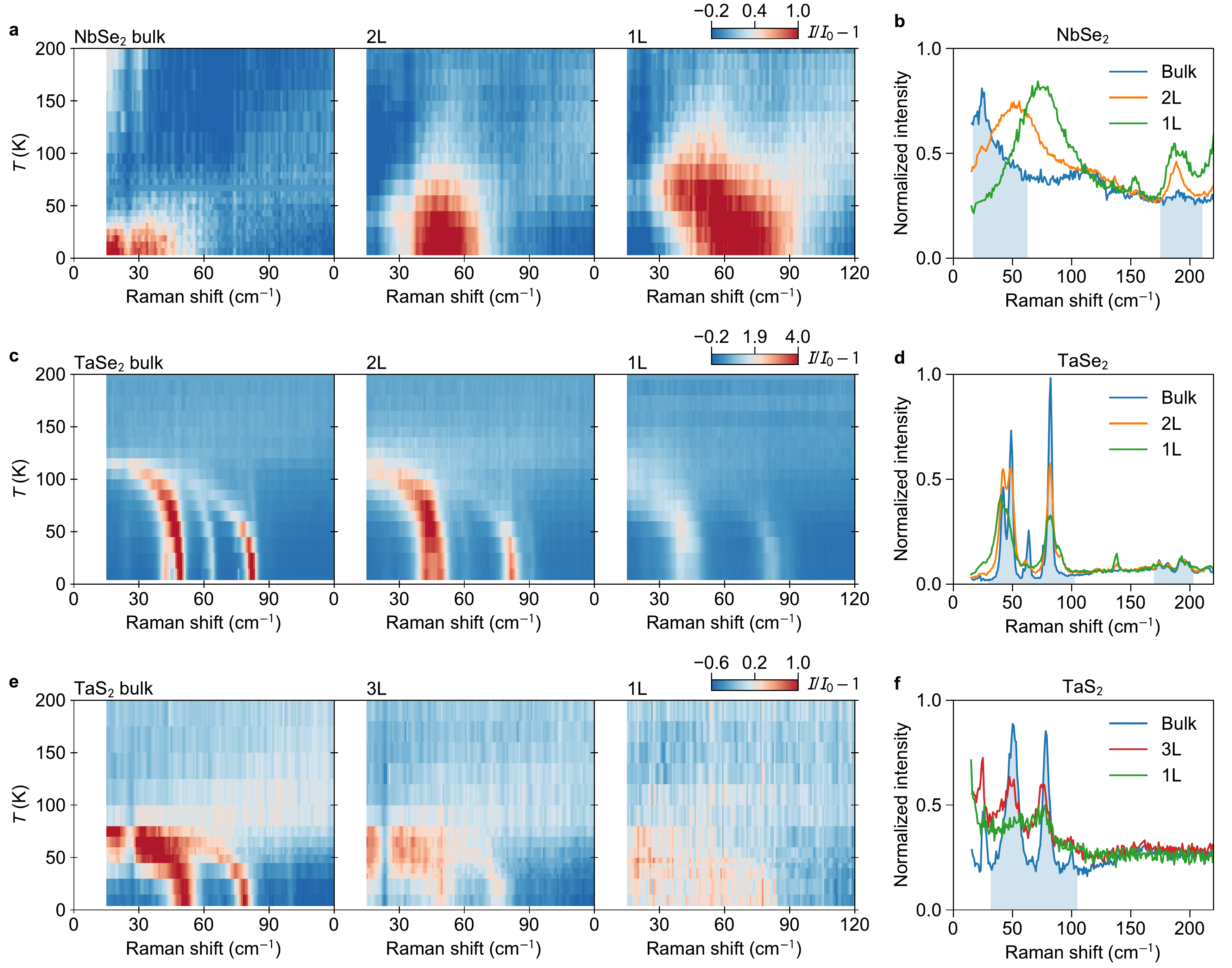}   
\caption{\textbf{Raman signature of CDW transitions in NbSe$_2$, TaSe$_2$, and TaS$_2$.} \textbf{a}, \textbf{c}, \textbf{e}, Temperature dependent Raman scattering intensity maps for the three compounds with different thickness, all collected in the collinear polarization configuration. To compare different samples, each set of raw data $I$ are normalized to a corresponding high-temperature spectrum $I_0$ far above $T_\mathrm{CDW}$, and unity is subtracted from the ratio to yield $I/I_0 - 1$. \textbf{b}, \textbf{d}, \textbf{f}, Comparison of the Raman scattering spectra at 4 K for samples of different thickness. In each panel, the spectra are normalized to match their background intensity between 150–220 cm$^{-1}$.} 
\label{Fig3}
\end{figure*}

\begin{figure*}
\includegraphics[scale=0.64]{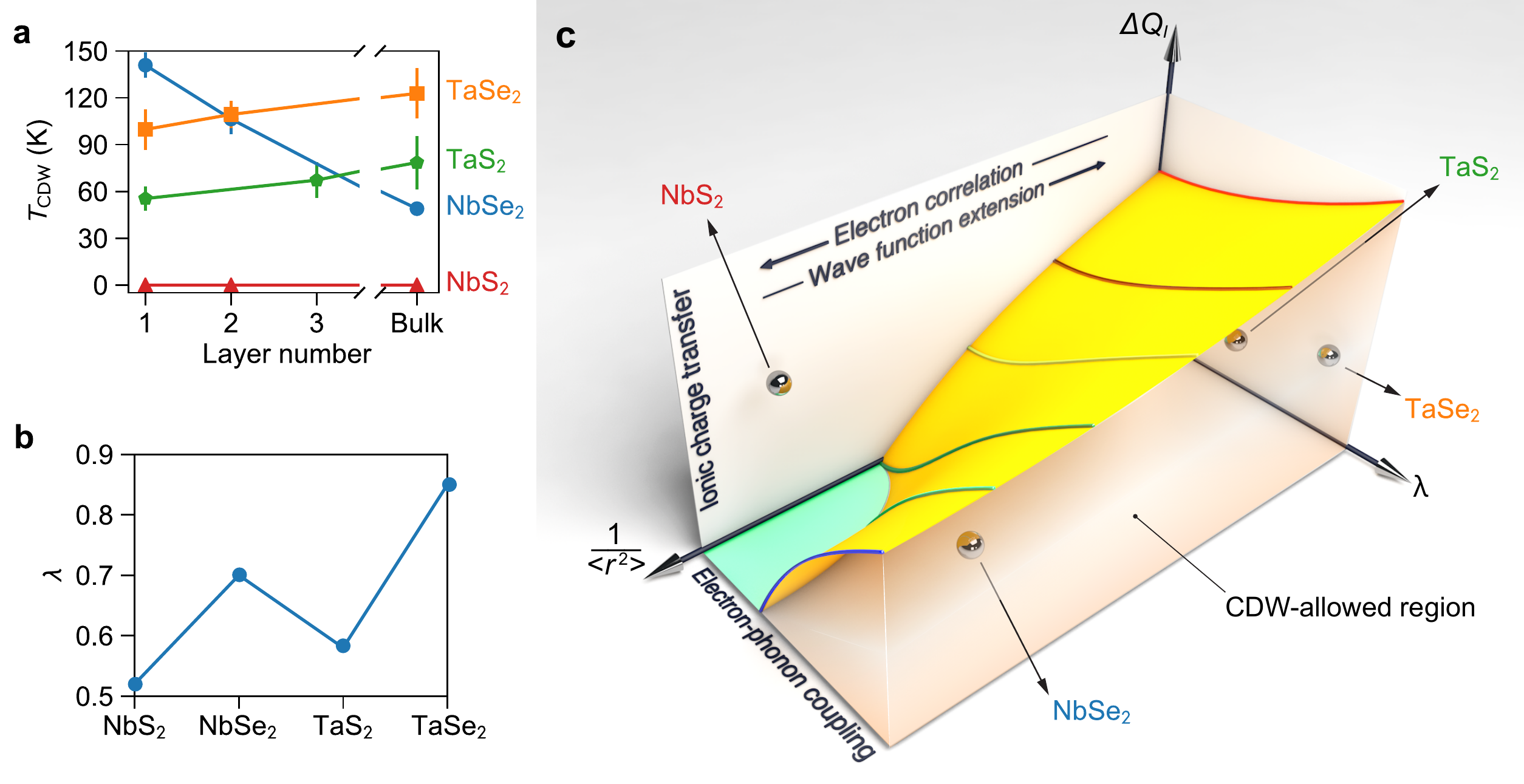}   
\caption{\textbf{Thickness dependence of $T_{\mathrm{CDW}}$ for 2\textit{H-MX}$_2$ (\textit{M}$=$Nb, Ta and \textit{X}$=$S, Se). } \textbf{a}, Layer-number dependence of $T_{\mathrm{CDW}}$ for all compounds. \textbf{b}, The calculated values of electron-phonon coupling constant $\lambda$ for monolayer 1\textit{H-MX}$_2$. \textbf{c}, Schematic illustration of the possible phase diagram describing the CDW response in a layered material in terms of ionic charge transfer $\Delta Q_I$, electron-phonon coupling constant $\lambda$ and the spatial extension of electronic wave functions $1/\langle r^2\rangle$.} 
\label{Fig4}
\end{figure*}

\end{document}